\documentclass[a4paper]{aa}  

\usepackage{graphicx}
\usepackage{txfonts}
\usepackage{lscape}
\usepackage{tabularx}
\usepackage{adjustbox}
\usepackage{hyperref}
\usepackage{comment}
\usepackage{stfloats}
\begin{document} 
\title{Chemical complexity in star formation induced by stellar feedback: cores shock-formed by the supernova remnant W44.}

\author{G. Cosentino$^{1}$\thanks{E-mail:cosentino@iram.fr}
I. Jim\'{e}nez-Serra$^{2}$,
F. Fontani$^{3,4,5}$,
P. Gorai$^{6,7}$,
C.-Y. Law$^{3}$,
J. C. Tan$^{8,9}$,
R. Fedriani$^{10}$,
A. T. Barnes$^{11}$,
P. Caselli$^{4}$,
S. Viti$^{12}$,
J. D. Henshaw$^{13}$}

\authorrunning{Cosentino et al.}
    \titlerunning{Chemical complexity in star formation induced by stellar feedback}
    \institute{Institut de Radioastronomie Millimétrique, 300 Rue de la Piscine, 38400 Saint-Martin-d’Hères, France
    \and Centro de Astrobiolog\'{i}a (CSIC/INTA), Ctra. de Torrej\'on a Ajalvir km 4, Madrid, Spain
    \and INAF  Osservatorio Astronomico di Arcetri, Largo E. Fermi 5, 50125 Florence, Italy
    \and Max Planck Institute for Extraterrestrial Physics, Giessenbachstrasse 1, 85748 Garching bei M\"{u}nchen, Germany
    \and Laboratory for the study of the Universe and eXtreme phenomena (LUX), Observatoire de Paris, 5, place Jules Janssen, 92195 Meudon, France
    \and Rosseland Centre for Solar Physics, University of Oslo, PO Box 1029 Blindern, 0315 Oslo, Norway
    \and Institute of Theoretical Astrophysics, University of Oslo, PO Box 1029 Blindern, 0315 Oslo, Norway
    \and Department of Space, Earth and Environment, Chalmers University of Technology, SE-412 96 Gothenburg, Sweden
    \and Department of Astronomy, University of Virginia, 530 McCormick Road Charlottesville, 22904-4325 USA
    \and Instituto de Astrof\'isica de Andaluc\'ia, CSIC, Glorieta de la Astronom\'ia s/n, E-18008 Granada, Spain
    \and European Southern Observatory (ESO), Karl-Schwarzschild-Straße 2, 85748 Garching bei M\"unchen, Germany
    \and Leiden Observatory, Leiden University, PO Box 9513, 2300 RA Leiden, The Netherlands
    \and Astrophysics Research Institute, Liverpool John Moores University, 146 Brownlow Hill, Liverpool L3 5RF, UK}

   \date{Received ---; accepted ---}

   \date{Received September 15, 1996; accepted March 16, 1997}

% \abstract{}{}{}{}{} 
% 5 {} token are mandatory
 
  \abstract
   {Low-velocity shocks from Supernova Remnants (SNRs) may set the physical and chemical conditions of star formation in molecular clouds. Recent evidence suggests that even the Sun might have formed through this process. However, the chemical conditions of shock-induced star forming regions remain poorly constrained.}
   {We study the chemical complexity of a shock-impacted clump, with potential to yield star formation, named the Clump, and located at the interface between the SNR W44 and the infrared dark cloud G034.77-00.55. We test whether the Clump has chemical properties consistent with those observed in star forming regions unaffected by SNRs.}
   {We use high-sensitivity, broad spectral surveys at 3 and 7 mm obtained with the 30m antenna at the Instituto de Radioastronomia Millímetrica and the 40 m antenna at the Yebes observatory, to identify D-bearing molecules and complex organic molecules (COMs) toward the Clump. For all species, we estimate molecular abundances and compare them with those observed across star forming regions at different evolutionary stages and masses, as well as comets.}
   {We detect multiple deuterated molecules (DCO$^+$, DNC, DCN, CH$_2$DOH) and COMs (CH$_3$OH, CH$_3$CHO, CH$_3$CCH, CH$_3$CN, CH$_3$SH) with excitation temperatures of 5–13 K. To the best of our knowledge, this is the first detection of COMs toward a site of SNR–cloud interaction. The derived D/H ratios ($\sim$0.01–0.04) and COM abundances are consistent with those reported toward typical low-mass starless cores and comparable to cometary values. The overall level of chemical complexity is relatively low, in line with an early evolutionary stage.}
   {We suggest that the Clump is a early stage shock-induced low-mass star forming region, not yet protostellar. We speculate that SNR-driven shocks may set the physical and chemical conditions to form stars. The resulting chemical budget may be preserved along the formation process of a planetary system, being finally incorporated into planetesimals and cometesimals.}


   \maketitle
%
%-------------------------------------------------------------------

\section{Introduction}
\begin{figure*}
    \centering
    \includegraphics[width=\linewidth,trim=1.5cm 0cm 2.5cm 0cm, clip=True]{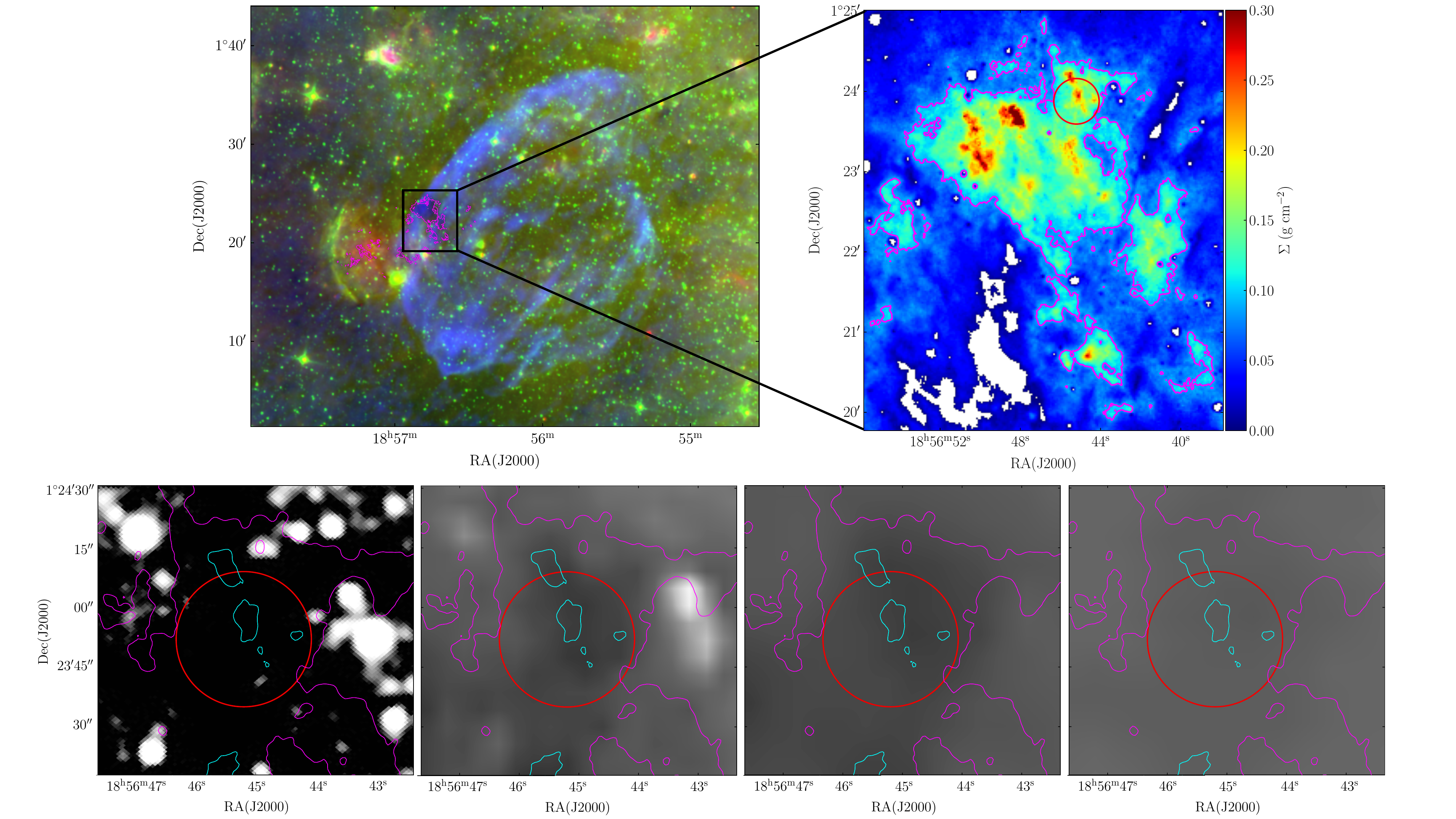}
    \caption{\textit{Top Left:} Three colour image of G34.77. Red is 24 $\mu$m emission from the MIPSGAL survey \citep{carey2009}, green is 8 $\mu$m emission from the GLIMPSE survey \citep{churchwell2009} and blue is 1 GHz continuum emission from the THOR survey \citep{beuther2016}. \textit{Top Right:} Mass surface density map \citep[color scale;][]{kainulainen2013} with the red circle indicating the location of the Clump. \textit{Bottom panels:} Zoom-in views of the Clump at 2 $\mu$m \citep[2MASS;][]{Skrutskie2006}, 8 $\mu$m \citep{carey2009}, 24 $\mu$m \citep{churchwell2009} and 70 $\mu$m \citep[Hi-GAL][]{Molinari2010} emission. The cyan contours correspond to a mass surface density value of 0.3 g cm$^{-2}$. In all panels, the magenta contour corresponds to a mass surface density of 0.1 g cm$^{-2}$ and highlights the cloud outskirts. The red circle size corresponds to 44$^{\prime\prime}$, i.e. the largest beam size in the Yebes-40m observations.}
    \label{fig:collage}
\end{figure*}

Supernova Remnants (SNRs) are among the most energetic \citep{2011IAUS..270..247B} and long-lasting \citep{leitherer1999,agertz2013} sources of feedback that affect the Interstellar Medium (ISM). Relatively slow shocks (a few tens of km s$^{-1}$) driven by late stages expanding SNR shells can sweep up the surrounding material, increasing its density \citep{inoueFukui2018, cosentino2019,cosentino2022,sano2020,sano2023} and enhancing its star formation potential \citep{barnes2023,khullar2024,Beattie2025}. The resulting high post‑shock densities enable rapid gas cooling and enhanced chemical reactions \citep{Fragile2003}, which can in turn trigger gravitational collapse even of initially stable clouds \citep{Li2014}. Recent simulations further suggest that shocks can funnel gas into filaments, setting the conditions under which gravity can drive star formation \citep{Federrath2012,Appel2023}.\\ There is also growing evidence that even our own Sun may have formed within a massive star-forming region shaped by stellar feedback \citep{Adams2010,Gounelle2012,Pfalzner2015}, where at least one nearby supernova played a pivotal role \citep{Boss2013,Young2014,Parker2023}. According to this scenario, a supernova shock impacting the proto-solar core may have triggered its collapse, while simultaneously enriching the material with radioactive and chemically processed ejecta \citep[e.g. ][]{BossKeiser2010}. The presence of short-lived radionuclides, such as $^{26}$Al and $^{60}$Fe, in primitive meteorites supports this scenario \citep{OuelletteDeschHester2007,OuelletteDeschHester2009,DeschYoung2022}. It has also been suggested that the Sun, as other Sun-like stars, may have emerged within complex hub–filament systems, where star formation is promoted by filament mergers induced by both early feedback in the form of H{\small II} regions and later-stage shocks from SNRs \citep{Arzoumanian2023}.\\

\noindent
Studying chemical complexity in dense cores within molecular clouds that are affected by the interaction with SNRs offers a unique opportunity to probe this scenario. However, direct observations of these environments remain scarce. Toward the SNRs W41 \citep{Hogge2019} and IC443 \citep{Reach2024}, indirect evidence of cores has been reported, but these may have pre-existed the shock and be negatively affected by it i.e., the shock is dispersing them, making it unlikely that they will collapse to form stars \citep{Zhou2021,cosentino2022,Reach2024}. More recently, \cite{cosentino2025a} have identified signatures of early-stage star formation towards the Infrared Dark Cloud (IRDC) G034.77-00.55 (thereafter G034.77), shock-interacting with the nearby SNR W44. W44 is a 20k years old, core-collapse, SNR located 2.9 kpc away \citep{Lee2020} and driving a relatively slow shock \citep[$\sim$15-20 km s$^{-1}$;][]{sashida2013,cosentino2019} into the western edge of G034.77. The author used high-angular resolution N$_2$H$^+$(1-0) ALMA images to reveal the presence of multiple cores at the shock interface. Several factors support the scenario that the cores formation may have been triggered by the shock passage. The density of the post-shocked gas ($\geq$10$^6$ cm$^{-3}$), traced by C$^{18}$O in \cite{cosentino2019} and N$_2$H$^+$ in \cite{cosentino2025a}, has been reported to be consistent with that typically found in star forming regions \citep{cosentino2019}. This dense post-shocked gas is significantly blue-shifted \citep[39-42 km s$^{-1}$][]{cosentino2019} with respect to the main cloud's velocity \citep[43-45 km s$^{-1}$][]{rathborne2006,hernandez2015}. The cores identified by \cite{cosentino2025a} show densities and kinematics consistent with that of the post-shocked gas. Moreover, the cores are distributed along two filaments whose orientation is consistent with that of the shock front. In \cite{cosentino2023}, the D/H ratio toward the post-shocked material was reported to be enhanced with respect to the inner cloud and comparable to that of other early stage star forming regions. Finally, the presence of N$_2$D$^+$ emission in the post-shocked gas suggests that this is likely dense and cold material, i.e. it has high potential to harbour star formation. Hence, the kinematic and morphological link between the post-shocked gas and the cores, as well as their similar physical conditions, suggests that the cores formation might have been enabled by the shock passage. The shock is therefore responsible for setting physical conditions within the cores that are similar to those observed in star forming regions.\\ The majority of the cores appear gravitationally unstable considering thermal and turbulent support. In the 1.2 mm maps obtained by \cite{rathborne2006} with the IRAM-30m MAMBO II instrument (11$^{\prime\prime}$ or 0.15 pc) and in the 3 mm ALMA images (3$^{\prime\prime}$ or 0.04 pc) reported by \cite{barnes2021} and \cite{cosentino2019} no continuum emission is found toward the cores. Finally, the cores are not associated with point-like sources at 8, 24 and/or 70 $\mu$m \citep{paron2009,barnes2016}, suggesting that no central protostars are present toward the cores. In light of all this, \cite{cosentino2025a} have suggested that shocks driven by the SNR W44 may have impacted the cloud, enhanced the gas density and triggered the formation of the identified cores. Hence, the cores may be regions of shock-induced star-formation.\\ 
\noindent
Across the region of G034.77 where the core population is located, the pointing previously identified as Clump \citep{cosentino2023} has the highest star formation potential \citep{cosentino2025a}. It shows high mass surface density $\Sigma >$0.1 g cm$^{-2}$ and a D/H ratio of $\sim$ 0.1, measured as the ratio between the N$_2$D$^+$ and N$_2$H$^+$ column densities obtained with single-pointing observations. This is consistent with values typically measured in both low-mass \citep{crapsi2005,friesen2013,cheng2021} and high-mass \citep{fontani2006,fontani2011,kong2016} starless and pre-stellar cores. Therefore, the Clump has the potential to be a unique laboratory for investigating the chemical complexity of prestellar cores whose formation might have been induced by stellar feedback.\\

\noindent
In this work, we investigate the Clump chemical inventory, also in relation with that of Galactic star forming regions at different evolutionary stages, as well as meteorites and comets, likely carriers of the pristine proto-solar environment. In particular, we measure deuterium fractionation in multiple species and assess the presence of complex organic molecules (COMs). D/H ratios derived from different tracers are valuable complementary indicators of a core’s evolutionary stage \citep{fontani2006,caselli2008,Ceccarelli2014,bianchi2017,sabatini2019,bovino2021,colzi2022}. For instance, deuterium fractionation measured from N$_2$D$^+$ and N$_2$H$^+$ increases during the starless/prestellar phase and decreases during the proto-stellar phase \citep{crapsi2005,emprechtinger2009}. On the other hand, deuteration measured from HNC and HCO$^+$ have less steep decline \citep{fontani2014,gerner2015}. For instance, \cite{fontani2015} found that the DNC/HNC column density ratio is less sensitive than the N$_2$D$^+$/N$_2$H$^+$ ratio across different stages of high-mass star formation. \\ COMs have been detected in a wide variety of environments \citep[see][for a review]{ceccarelli2022}. Traditionally associated with hot cores \citep[e.g.][]{Brown1975,Blake1987} and corinos \citep{Cazaux2003, Bianchi2019, martin2021}, COMs have also been observed in starless and pre-stellar cores \citep[e.g.][]{jimenezserra2016,scibelli2021,megias2023,Scibelli2024}, dark clouds \citep[e.g.][]{Agundez2015,Marcelino2020,Cernicharo2021}, translucent clouds \citep{Thiel2017}, and protoplanetary disks \citep[e.g.][]{Walsh2014}. Their abundance evolution during the star-formation process has been investigated in recent studies \citep{Coletta2020,Bhat2023}, which indicates an increasing trend from the starless/pre-stellar phase to the protostellar phase. \\ 
The paper is organised as follows. In Sect.~\ref{sec:ObsAndData}, we describe data and observation strategy. In Sect.~\ref{sec:abundances}, we identify the molecular species of interest and measure their abundances. In Sect.~\ref{sec:results}, we present and discuss our results. Finally, in Sect.~\ref{sec:conclusions}, we present our conclusions. 

\section{Observations and Data}\label{sec:ObsAndData}
\subsection{IRAM-30m observations}
In this work, we use 3 mm and 7 mm single pointing observations obtained in March 2020 with the 30m antenna at the Instituto de Radioastronomia Millímetrica (IRAM; Pico Veleta, Spain) and in April 2024 with the 40m antenna at the Yebes Observatory (Castilla-La Mancha, Spain), respectively. Observations have been acquired toward the position previously identified as Clump by \cite{cosentino2023} and centered at RA(J2000)=18$^h$56$^m$45.2$^s$, Dec(J2000)=1$^d$23$^m$52$^s$) with off position RA(J2000)=18$^h$57$^m$01$^s$, Dec(J2000)=1$^d$22$^m$25$^s$. In Figure~\ref{fig:collage}, the location of the IRDC G034.77 with respect to the SNR W44 is shown (top left panel). The location of the Clump within the cloud is also indicated (top right panel), as well as a zoom-in view at multiple wavelengths from archival data (bottom panels). From Figure~\ref{fig:collage}, no IR signatures of deeply embedded protostars are found, confirming the results already presented by \cite{barnes2021,cosentino2025a}.\\ Both sets of observations were performed in dual polarisation and position-switching mode. For the IRAM-30m observations, we used the EMIR90 receiver, tuned at frequencies of 74.0 and 86.0 GHz, to cover a continuous frequency range of 71.9-102.3 GHz. We also used the FTS spectrometer with a frequency resolution of 200 kHz, corresponding to a velocity resolution between 0.7 and 0.8 km\,s$^{-1}$. Intensities were measured in units of antenna temperature, T$^{*}_{A}$, and converted into main-beam brightness temperature, $T_{\rm mb}$, using beam and forward efficiencies of 0.81 and 0.95, respectively. The angular resolution of the 30m antenna is between 34$^{\prime\prime}$ and 24$^{\prime\prime}$ in the 3 mm band. These correspond to a linear spatial resolution of 0.4-0.5 parsec at the distance of G34.77 i.e., 2.9 kpc \citep{Lee2020}.

\subsection{YEBES-30m observations}
The Yebes-40m observations at 7 mm were performed in Q-band, covering a continuous frequency range 31.3-49.5 GHz and using the FFT spectrometer with spectral resolution of 38 kHz ($\sim$ 0.25 km s$^{-1}$). In order to increase the signal-to-noise ratio, we smoothed the spectra in velocity, using the function {\sc smooth} in {\sc gildas} and obtaining final velocity resolution in the range 0.5-0.7 km s$^{-1}$. Intensities were converted from T$^{*}_{A}$ to T$_{\rm mb}$, using beam and forward efficiencies of 0.61 and 0.95, respectively. Observations performed with the 40m antenna have angular resolution between 36$^{\prime\prime}$ and 44$^{\prime\prime}$ i.e., 0.5-0.6 pc at the source distance. All these information are summarised in Table~\ref{tab:TelescopeSetUp}. The final spectra were produced using the {\sc CLASS} software within the {\sc gildas} package\footnote{See http://www.iram.fr/IRAMFR/GILDAS.} and have rms per channel between 7 and 5 mK for the IRAM-30m observations and 3-5 mK for the Yebes-40m observations.

\subsection{MIREX map}
For our analysis, we also employ the mass surface density, $\Sigma$, map obtained toward G034.77 by \cite{kainulainen2013} and shown in the top right panel of Figure~\ref{fig:collage}. In order to obtain this map, the authors extended the MIREX method previously developed by \cite{butlerTan2012}, combining multi-wavelengths Mid- and Near-Infrared Spitzer images of the cloud. The obtained mass surface density map has angular resolution of 2$^{\prime\prime}$, consistent with the resolution of the Spitzer-IRAC maps and the uncertainty is estimated to be $\sim$30\% \citep{kainulainen2013}. Since the MIREX method is not suitable to estimate mass surface densities toward sources that are bright at Mid-IR wavelengths, pixels toward which the method is not applicable appear as blank (white) in the map. For a more detailed description of the MIREX method we refer to \cite{butlerTan2009,butlerTan2012} and \cite{kainulainen2013}.

\begin{table}
    \centering
    \caption{Telescope, frequency ranges, rms noise (T$_{\rm{mb}}$), velocity resolution and beam-aperture of the observations presented in this work.}
    \begin{tabular}{ccccc}
    \hline
    \hline
    Telescope & Frequency & rms & $\delta v$ & Beam \\
    & (GHz)     & (K)   & (km s$^{-1}$) & ($^{\prime\prime}$) \\
    \hline
    Yebes-40m & 31.10 - 49.50 & 3 - 7 & 0.7 - 0.5 & 44-36\\
    IRAM-30m  & 71.70 - 87.10 & 7 - 5 & 1.0 - 1.4 & 34-28\\
    IRAM-30m  & 87.38 - 102.6 & 6 - 6 & 1.3 - 1.1 & 28-24\\
    \hline
    \end{tabular}
    \label{tab:TelescopeSetUp}
\end{table}

\section{Method}\label{sec:abundances}
\subsection{Line identification}
\begin{figure*}
    \centering
    \includegraphics[width=\linewidth,trim= 3cm 0.8cm 3.5cm 1cm, clip=True]{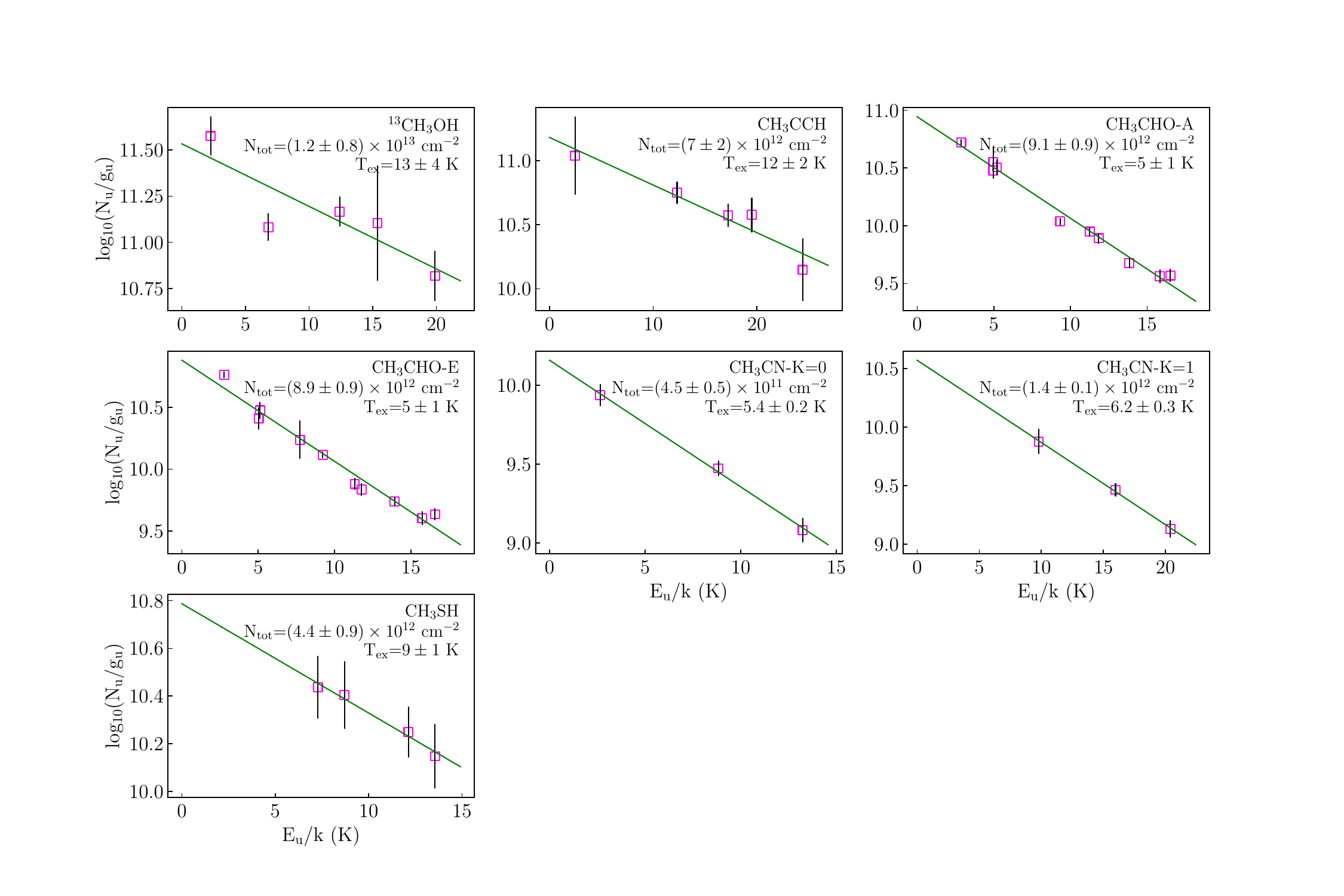}
    \caption{Rotational diagrams obtained using {\sc madcuba} for $^{13}$CH$_3$OH, CH$_3$CCH, CH$_3$CHO, CH$_3$CN and CH$_3$SH. Magenta empty squares indicate the data points, while the green lines show the best fitting found by {\sc madcuba}. In all panels, the corresponding species is indicated together with the inferred excitation temperature, $T_{\mathrm{ex}}$ and total column density, $N_{\mathrm{tot}}$.}
    \label{fig:RotDiagrams}
\end{figure*}

The first step in our analysis is to identify species of interest in the large bandwidth spectra. For this, we have used the tool \textit{Spectral Line Identification and Modelling}, {\sc slim} within the {\sc madcuba} software\footnote{Madrid Data Cube Analysis on ImageJ is a software developed at the Center of Astrobiology (CAB) in Madrid; http://cab.intacsic.es/madcuba/Portada.html} \citep{martin2019}. {\sc madcuba} is a stand-alone, interactive tool specifically tailored for the analysis of spectra and spectral cubes. It allows to perform line identification and line modelling in Local-Thermodynamic-Equilibrium (LTE) conditions. Using the spectroscopic parameters available in the Cologne Database for Molecular Spectroscopy\footnote{https://cdms.astro.uni-koeln.de/} \citep[CDMS;][]{endres2016} and the Jet Propulsion Laboratory\footnote{https://spec.jpl.nasa.gov/} \citep[JPL;][]{Pickett1998}, we have used {\sc slim} to identify D-bearing species (and their H-bearing counterparts) and COMs within our spectra. We consider as detection all those transitions for which the integrated intensity is at least a factor of 3 larger than the integrated noise\footnote{We consider a linewidth of 3 km s$^{-1}$, typical of the detected species, and estimate the integrated noise as A$_{\mathrm{rms}}$=rms$\times\delta$v$\times \sqrt{N_{ch}}$, where rms is the spectral noise, $\delta v$ is the velocity resolution and $N_{ch}\sim$4-6 is the number of channel within the 3 km s$^{-1}$ linewidth.}. In Table~\ref{tab:LineList}, we report the transitions, frequencies and spectral parameters as well as the integrated signal-to-noise ratio for each detected species. Among the identified deuterated species we report DCO$^+$, DNC and DCN. Within the COMs group, we identify CH$_3$OH, CH$_3$CHO, CH$_3$CCH, CH$_3$SH and CH$_3$CN. Moreover, we report detection of two transitions of singly-deuterated Methanol, CH$_2$DOH. We note that bright species such as HCO$^+$, HNC, HCN and CH$_3$OH, counterparts to the identified D-bearing species, are detected but they are very bright and strongly suffer from optical depth effects. Hence, for further analysis, we consider their $^{13}$C-bearing counterparts, which are most optically thin but still significantly bright (S/N$\geq$50). This is confirmed by {\sc MADCUBA}, which indicated for all identified species optical depth $\tau\leq$0.03. Within {\sc madcuba}, we also use the function {\sc autofit} to obtain the best fit to the species spectral profiles. For each species, the function assumes all transitions to have the same central velocity and linewidth and return the LTE model that best reproduces the line profile of all transitions simultaneously. The obtained best fit values of central velocity, linewidth and integrated flux are listed in Table~\ref{tab:LineList}. 

\begin{table*}
\centering
\caption{Molecular species, transition, frequency, A$_{ul}$, E$_{\mathrm{low}}$, integrated intensity, central velocity, linewidth and signal-to-noise ratio of all the detected transition of interest.}
\begin{tabular}{llccccccc}
\hline
\hline
Molecule & Transition & Frequency & log$_{10}$($A_{\rm{ul}}$) & E$_{\rm{low}}$ & Area & $v_0$ & $\Delta v$ & S/N\\
         &$J_{k_a,k_c}$-$J^{\prime}$$_{k^{\prime}_a,k^{\prime}_c}$  & (MHz)  & (s$^{-1}$) & (K) & (K km s$^{-1}$) & (km s$^{-1}$) & (km s$^{-1}$) & \\
\hline
H$^{13}$CO$^+$  &1-0                 &86754.288 &-2.2835 &0.00 &1.232$\pm$0.007 &41.9$\pm$0.1 &3.2$\pm$0.1 &246\\
DCO$^+$         &1-0                 &72039.241 &-3.4952 &0.00 &0.177$\pm$0.009	&41.9$\pm$0.1 &3.3$\pm$0.1 &198\\
HN$^{13}$C      &1-0                 &87090.850 &-2.5950 &0.00 &0.456$\pm$0.04 &41.8$\pm$0.1 &2.4$\pm$0.3 &17\\
DNC             &1-0                 &76305.727  & -2.661 & 0.00 &0.641$\pm$0.005 &41.9$\pm$0.1 &3.3$\pm$0.1 &100\\
H$^{13}$CN      &1-0 F=1-1        &86338.767   & -2.9964 & 0.00 &0.201$\pm$0.003 &41.9$\pm$0.1 &3.6$\pm$0.1 &30\\
                &1-0 F=2-1        &86340.184   & -2.7746 & 0.00 &0.335$\pm$0.003 &41.9$\pm$0.1 &3.6$\pm$0.1 &50\\
                &1-0 F=0-1        &86342.274   & -3.4735 & 0.00 &0.067$\pm$0.003 &41.9$\pm$0.1 &3.6$\pm$0.1 &10\\
DCN             &1-0 F=1-1        &72413.484   & -3.2249 & 0.00 &0.088$\pm$0.005 &41.7$\pm$0.1 &3.3$\pm$0.3 &10\\
                &1-0 F=2-1        &72414.905  & -3.0031 & 0.00 &0.147$\pm$0.005 &41.7$\pm$0.1  &3.3$\pm$0.3 &18\\
                &1-0 F=0-1        &72417.030   & -3.7020 & 0.00 &0.030$\pm$0.005 &41.7$\pm$0.1  &3.3$\pm$0.3 &5\\
\hline
$^{13}$CH$_3$OH &1$_{0,1}$-0$_{0,0}$   &47205.210   &-5.9005 &0.00  &0.80$\pm$0.01 &41.5$\pm$0.3 &3.7$\pm$0.6 &133\\
                &2$_{-1,2}$-1$_{-1,1}$ &94405.163   &-5.1352 &5.47  &0.71$\pm$0.01 &41.5$\pm$0.3 &3.7$\pm$0.6 &156\\
                &2$_{0,2}$-1$_{0,1}$   &94407.129   &-5.0024 &1.57  &0.85$\pm$0.01 &41.5$\pm$0.3 &3.7$\pm$0.6 &189\\
                &2$_{1,2}$-1$_{1,1}$   &94411.016   &-5.0215 &10.69 &0.38$\pm$0.01 &41.5$\pm$0.3 &3.7$\pm$0.6 &83\\
CH$_2$DOH       &1$_{0,1}$-0$_{0,0}$   &44713.190 & -6.2750 & 0.00 &0.060$\pm$0.004 &41.0$\pm$0.1 &3.9$\pm$0.3 &13\\
                &2$_{0,2}$-1$_{0,1}$   &89407.817 & -5.3767 & 1.49 &0.160$\pm$0.004 &41.0$\pm$0.1 &3.9$\pm$0.3 &21\\
CH$_3$CHO       &2$_{1,2}$-1$_{1,1}$ (A) &37464.204 & -5.6959 & 2.20 &0.72$\pm$0.005 &42.0$\pm$0.1 &3.6$\pm$0.1 &193\\
                &2$_{1,2}$-1$_{1,1}$ (E) &37686.932 & -5.7200 & 2.24 &0.08$\pm$0.005 &42.0$\pm$0.1 &3.6$\pm$0.1 &20\\
                &2$_{0,2}$-1$_{0,1}$ (A) &38506.035 & -5.5438 & 0.151 &0.18$\pm$0.005 &42.0$\pm$0.1 &3.6$\pm$0.1 &45\\
                &2$_{0,2}$-1$_{0,1}$ (E) &38512.079 & -5.5438 & 0.154 &0.22$\pm$0.005 &42.0$\pm$0.1 &3.6$\pm$0.1 &59\\
                &2$_{1,1}$-1$_{1,0}$ (A) &39362.537 & -5.6825 & 2.30 &0.09$\pm$0.005 &42.0$\pm$0.1 &3.6$\pm$0.1 &22\\
                &2$_{1,1}$-1$_{1,0}$ (E) &39594.289 & -5.648 & 2.24  &0.10$\pm$0.005 &42.0$\pm$0.1 &3.6$\pm$0.1 &25\\
                &4$_{1,4}$-3$_{1,3}$ (A) &74891.677 & -4.7041 & 5.33 &0.17$\pm$0.005 &42.0$\pm$0.1 &3.6$\pm$0.1 &18\\
                &4$_{1,4}$-3$_{1,3}$ (E) &74924.134 & -4.7046 & 5.38 &0.09$\pm$0.005 &42.0$\pm$0.1 &3.6$\pm$0.1 &7\\
                &4$_{0,4}$-3$_{0,3}$ (A) &76866.436 & -4.6506 & 3.92 &0.18$\pm$0.005 &42.0$\pm$0.1 &3.6$\pm$0.1 &16\\
                &4$_{0,4}$-3$_{0,3}$ (E) &76878.953 & -4.6506 & 3.85 &0.23$\pm$0.005 &42.0$\pm$0.1 &3.6$\pm$0.1 &28\\
                &4$_{1,3}$-3$_{1,2}$ (A) &79099.313 & -4.6581 & 5.59 &0.13$\pm$0.005 &42.0$\pm$0.1 &3.6$\pm$0.1 &21\\
                &4$_{1,3}$-3$_{1,2}$ (E)&79150.166 & -4.6567 & 5.54 &0.12$\pm$0.005 &42.0$\pm$0.1 &3.6$\pm$0.1 &17\\
                &2$_{1,2}$-1$_{0,1}$ (A) &84219.749 & -5.7339 & 0.64 &0.05$\pm$0.005 &42.0$\pm$0.1 &3.6$\pm$0.1 &8\\
                &5$_{1,5}$-4$_{1,4}$ (E) &93580.909 & -4.4093 & 7.82 &0.09$\pm$0.005 &42.0$\pm$0.1 &3.6$\pm$0.1 &9\\
                &5$_{1,5}$-4$_{1,4}$ (A) &93595.25  & -4.4093 & 7.87 &0.07$\pm$0.005 &42.0$\pm$0.1 &3.6$\pm$0.1 &6\\
                &5$_{0,5}$-4$_{0,4}$ (E) &95947.437 & -4.3672 & 6.48 &0.14$\pm$0.005 &42.0$\pm$0.1 &3.6$\pm$0.1 &13\\
                &5$_{0,5}$-4$_{0,4}$ (A) &95963.459 & -4.3672 & 6.42 &0.12$\pm$0.005 &42.0$\pm$0.1 &3.6$\pm$0.1 &11\\
                &5$_{1,4}$-4$_{1,3}$ (E) &98863.314 & -4.3627 & 8.23 &0.11$\pm$0.005 &42.0$\pm$0.1 &3.6$\pm$0.1 &10\\
                &5$_{1,4}$-4$_{1,3}$ (A) &98900.945 & -4.3622 & 8.18 &0.10$\pm$0.005 &42.0$\pm$0.1 &3.6$\pm$0.1 &9\\
                &3$_{1,3}$-2$_{0,2}$ (E) &101343.441 & -5.4569& 1.99 &0.04$\pm$0.005 &42.0$\pm$0.1 &3.6$\pm$0.1 &4\\
CH$_3$CCH       &2$_1$-1$_1$ &34182.755 & -5.9303 &5.58  &0.040$\pm$0.005 &41.5$\pm$0.06 &2.9$\pm$0.3 &5\\
                &2$_0$-1$_0$ &34183.420 & -5.7949 &0.57  &0.020$\pm$0.005 &41.5$\pm$0.06 &2.9$\pm$0.3 &5\\
                &5$_1$-4$_1$ &85455.622 & -4.6417 &10.70 &0.050$\pm$0.005 &41.5$\pm$0.06 &2.9$\pm$0.3 &5 \\
                &5$_0$-4$_0$ &85457.272 & -4.6135 & 5.70 &0.090$\pm$0.005 &41.5$\pm$0.06 &2.9$\pm$0.3 &9\\
                &6$_1$-5$_1$ &102546.024 & -4.4052 &13.55 &0.04$\pm$0.005 &41.5$\pm$0.06 &2.9$\pm$0.3 &3\\
                &6$_0$-5$_0$ &102547.984 & -4.3825 & 8.55 &0.08$\pm$0.005 &41.5$\pm$0.06 &2.9$\pm$0.3 &7\\
CH$_3$SH        &3$_{0,3}$-2$_{0,2}$ (A$^+$) &75862.889 & -5.0852 & 2.53 &0.050$\pm$0.005 &41.8$\pm$0.1 &2.8$\pm$0.3 &5\\
                &3$_{0,3}$-2$_{0,2}$ (E$^+$) &75864.422 & -5.0869 & 3.52 &0.040$\pm$0.005 &41.8$\pm$0.1 &2.8$\pm$0.3 &4\\
                &4$_{0,4}$-3$_{0,3}$ (A$^+$) &101139.150 & -4.7166 & 5.06 &0.050$\pm$0.005 &41.8$\pm$0.1 &2.8$\pm$0.3 &9\\
                &4$_{0,4}$-3$_{0,3}$ (E$^+$) &101139.654 & -4.7183 & 6.05 &0.050$\pm$0.005 &41.8$\pm$0.1 &2.8$\pm$0.3 &9\\
CH$_3$CN        &2$_1$-1$_1$ F=1-1 &36794.765 & -4.5607 & 5.58 &0.020$\pm$0.005  &41.4$\pm$0.1 &2.8$\pm$0.3 &5\\
                &2$_0$-1$_0$ F=2-1 &36795.475 & -4.4253 & 0.61 & 0.03$\pm$0.005 &41.4$\pm$0.1 &2.8$\pm$0.3 &5\\
                &4$_1$-3$_1$ F=4-4 &73588.799 & -3.5684 & 8.65 &0.082$\pm$0.005 &41.4$\pm$0.1 &2.8$\pm$0.3 &7\\
                &4$_0$-3$_0$ F=3-2 &73590.218 & -3.5299 & 3.68 &0.104$\pm$0.005 &41.4$\pm$0.1 &2.8$\pm$0.3 &8\\
                &5$_1$-4$_1$ F=6-5 &91985.314 & -3.2731 & 11.10 &0.118$\pm$0.005 &41.4$\pm$0.1 &2.8$\pm$0.3 &9\\
                &5$_0$-4$_0$ F=5-4 &91987.088 & -3.2450 & 6.14 &0.147$\pm$0.005 &41.4$\pm$0.1 &2.8$\pm$0.3  &10\\
\hline
\end{tabular}
\label{tab:LineList}
\end{table*}

\subsection{Molecular abundances}
For each detected molecule, we estimate the molecular abundance with respect to H$_2$ as:

\begin{equation}
    \chi \mathrm{(X)} = \frac{N\mathrm{(X)}}{N\mathrm{(H_2)}}
    \label{eq:Abundances}
\end{equation}

\noindent 
where $N$(X) and $N$(H$_2$) are the species and the molecular Hydrogen total column densities, respectively. In order to estimate $N$(H$_2$), we consider the mean $\Sigma$ value within the largest beam in our observations, and convert this to total column density, using a molecular Hydrogen mass of m(H$_2$)=3.67$\times$10$^{-24}$ g, as follows:

\begin{equation}
    {N\mathrm{(H_2)}} = \frac{\Sigma}{m\mathrm{(H_2)}}
\end{equation}

\noindent 
The employed molecular Hydrogen mass also include a 10\% contribution of Helium per Hydrogen atom. The derived molecular Hydrogen column density is $N$(H$_2$)=(3.8$\pm$1.1)$\times$10$^{22}$ cm$^{-2}$, where the uncertainty is due to the 30\% uncertainty in the $\Sigma$ value \citep{kainulainen2013}.\\

\noindent
In order to estimate the total column density, $N$(X), of all the detected molecules, we first analyse species for which more than three transitions (with significantly different excitation energies) have been detected. This is the case of $^{13}$CH$_3$OH, CH$_3$CHO, CH$_3$CCH and CH$_3$CN. For these species, we have used {\sc madcuba} to build rotational diagrams from the line integrated intensities and estimate the molecule excitation temperature, $T_{\mathrm{ex}}$, and total column density, $N$(X). The obtained rotational diagrams are shown in Figure~\ref{fig:RotDiagrams}. We obtain excitation temperatures in the range $T_{\mathrm{ex}}\sim$5-13 K and total column densities $N_{\mathrm{tot}}\sim$0.8-20$\times$10$^{12}$ cm$^{-2}$. We note that the $T_{\mathrm{ex}}$ obtained for $^{13}$CH$_3$OH and other COMs is consistent, within the uncertainty, with $T_{\mathrm{ex}}$=9 K estimated for CH$_3$OH in \cite{cosentino2018}, with the only exception is CH$_3$CHO, for which we report a lower $T_{\mathrm{ex}}\sim$5 K, in both the A and E forms.\\

\noindent 
For the remaining species listed in Table~\ref{tab:LineList}, we either detect only one rotational transition (H$^{13}$CO$^+$, DCO$^+$, HN$^{13}$C, DNC) or multiple transitions with very little energy difference, i.e. not sufficient to build rotational digrams (H$^{13}$CN, DCN and CH$_2$DOH). For these species, we fix the excitation temperature at both $T_{\mathrm{ex}}$=5 K and 13 K, i.e., the minimum and maximum $T_{\mathrm{ex}}$ found with the multi-transition approach, and use the {\sc autofit} function in {\sc slim} to estimate the total column density. Hence, we consider the best values as the mean between the two obtained column densities and associate an uncertainty as half the difference between the two values. The obtained values are then converted into abundances using Eq.~\ref{eq:Abundances}. Finally, for all the detected D-bearing species, we calculate the Deuterium fractionation (D/H ratio) as the ratio between the column densities of the deuterated species and its non deuterated counterpart. For the latter, when derived from the $^{13}$C-bearing isotopologue, we we assume isotopic ratio  $^{12}$C/$^{13}$C=49.8, as specifically calculated by \cite{zeng2017} for G034.77. We also note that, as described by \cite{Parise2004} and \cite{Manigand2019}, the D/H ratio from CH$_2$DOH and CH$_3$OH corresponds to 3 times the column density ratio between these species. The final value of column density, excitation temperature, molecular abundance and deuterium fractionation obtained with the above method are reported in Table~\ref{tab:molecularAbundances}, where we also list the COMs abundance relative to CH$_3$OH. The CH$_3$OH column density has been obtained from $N$($^{13}$CH$_3$OH), assuming $^{12}$C/$^{13}$C=49.8 \citep{zeng2017}. We estimate $N$(CH$_3$OH)=(6$\pm$4)$\times$10$^{14}$ cm$^{-2}$, a value consistent with those previously reported toward the massive cores G1, G2 and G3 within G034.77 \citep{cosentino2018}.\\

\noindent
We note that a filling factor = 1 was assumed in the analysis described above. This assumption was adopted because we do not know the actual source size, required to correct for the beam dilution effect. Moreover, the Clump most likely host multiple condensations \citep{cosentino2025a} and one should also account for the number of sources within the beam. Hence, since both pieces of information are uncertain, we do not correct for them and the COM abundance values reported here may be regarded as lower limits.

\begin{table*}
\centering
\caption{Molecular species, total column density, excitation temperature, molecular abundance with respect to H$_2$, Deuterium fractionation values and COM abundances with respect to CH$_3$OH obtained for all the species of interest.}
\begin{tabular}{lcccccc}
\hline
\hline
 Molecule        &   $N_{\mathrm{tot}}$ &   $T_{\mathrm{ex}}$ & $\chi$ &D/H &$N$(X)/$N$(CH$_3$OH)\\
                 & ($\times$10$^{12}$ cm$^{-2}$)  &   (K) & ($\times$10$^{-10}$) & \\
\hline
H$^{13}$CO$^+$  &1.61$\pm$0.02  &5-13           &0.4 $\pm$ 0.1      &$\cdots$ &$\cdots$\\
DCO$^+$         &2.78$\pm$0.04  &5-13           &0.7 $\pm$ 0.2      &0.04$\pm$0.01 &$\cdots$\\
HN$^{13}$C      &1.1$\pm$0.1    &5-13           &0.3 $\pm$ 0.1      &$\cdots$ &$\cdots$\\
DNC             &1.45$\pm$0.03  &5-13           &0.40 $\pm$ 0.03     &0.03$\pm$0.01 &$\cdots$\\
H$^{13}$CN      &1.3$\pm$0.02   &5-13           &0.3 $\pm$ 0.1      &$\cdots$ &$\cdots$\\
DCN             &0.76$\pm$0.05  &5-13           &0.20 $\pm$ 0.06    &0.012$\pm$0.006 &$\cdots$\\
$^{13}$CH$_3$OH &12$\pm$8       &13$\pm$4       &3$\pm$2            &$\cdots$ &$\cdots$\\
CH$_2$DOH       &19$\pm$1       &5-13           &4.5 $\pm$1.6       &0.09$\pm$0.06 &$\cdots$\\
CH$_3$CCH       &7$\pm$2        &12$\pm$2       &1.7$\pm$0.7        &$\cdots$ &0.012$\pm$0.008\\
CH$_3$CHO$^a$   &18$\pm$2       &5$\pm$1        &2.1$\pm$0.6        &$\cdots$ & 0.03$\pm$0.02\\
CH$_3$CN        &1.9$\pm$0.1    &5.4-6.2      &0.5$\pm$0.1        &$\cdots$ &0.003$\pm$0.002\\
CH$_3$SH        &4.4$\pm$0.9    &9$\pm$1        &1.1$\pm$0.4        &$\cdots$ & 0.007$\pm$0.005\\
\hline  
\end{tabular}
\tablefoot{$^{a}$ The reported CH$_3$CHO column density has been obtained as the sum of the A+E values.}
\label{tab:molecularAbundances}
\end{table*}

\section{Results and Discussion}\label{sec:results}
\subsection{D-bearing Species}\label{sec:deuterated}
\begin{figure}[!htpb]
    \centering
    \includegraphics[width=\linewidth, trim= 0cm 3.9cm 0cm 0.2cm, clip=True]{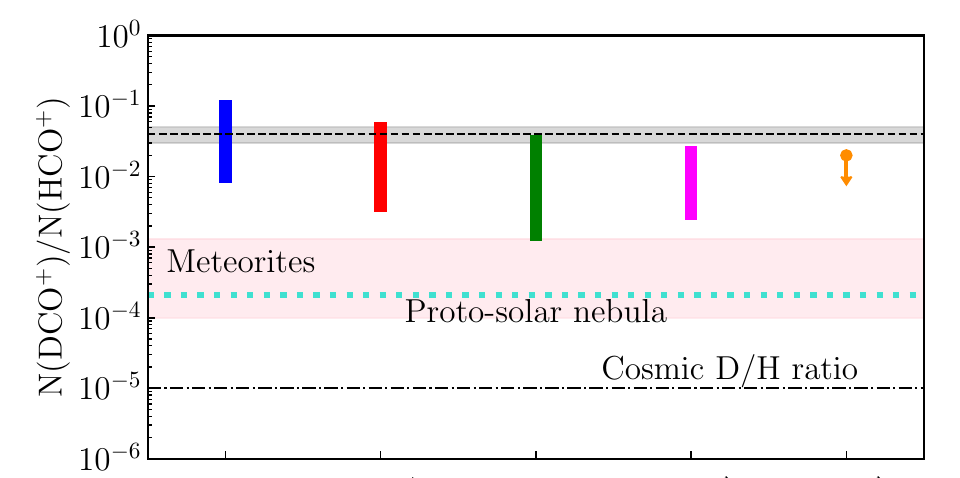}
    \includegraphics[width=\linewidth, trim= 0cm 3.9cm 0cm 0.2cm, clip=True]{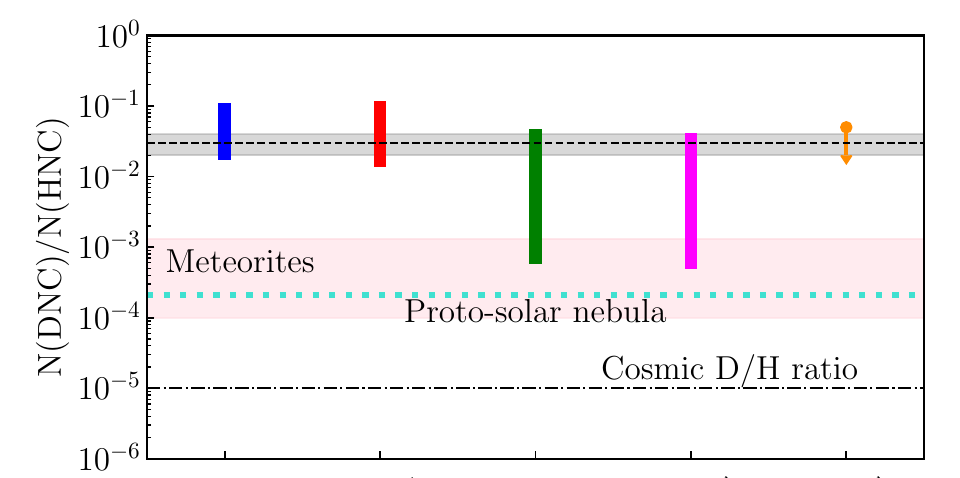}
    \includegraphics[width=\linewidth, trim= 0cm 3.9cm 0cm 0.2cm, clip=True]{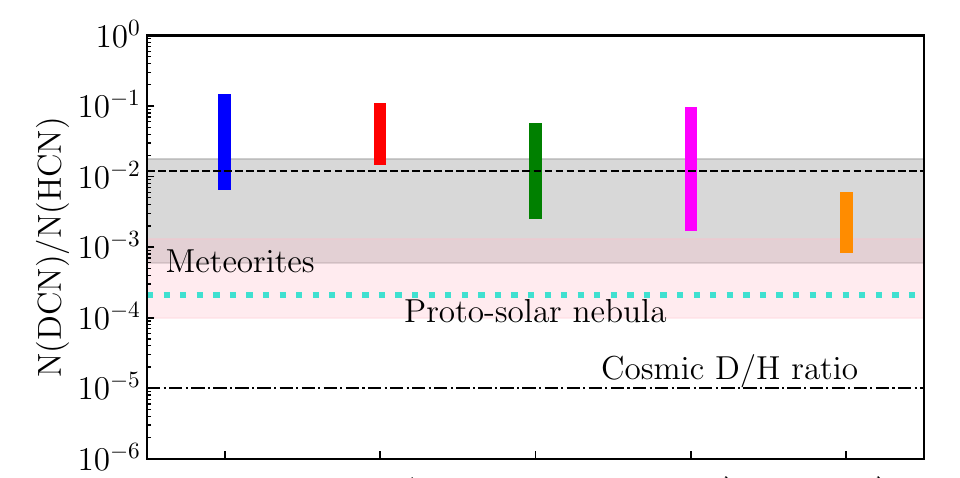}
    \includegraphics[width=\linewidth, trim= 0cm 0.5cm 0cm 0.2cm, clip=True]{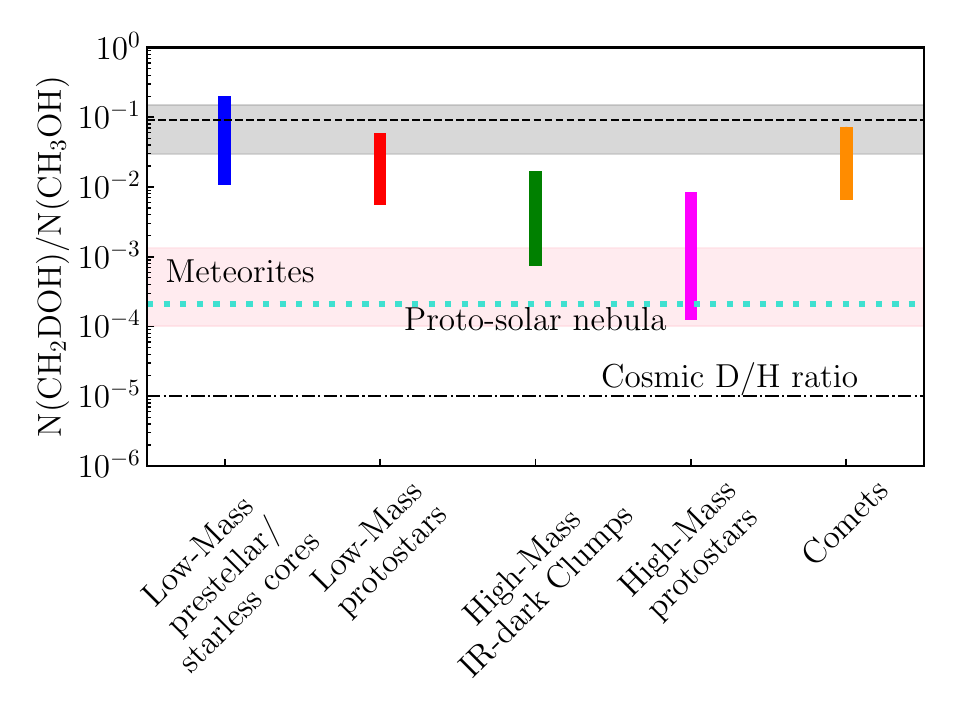}
    \caption{D/H ratios measured from DCO$^+$, DNC, DCN and CH$_2$DOH toward the Clump (dashed black lines and grey shadows) compared with literature values across low-mass starless/pre-stellar cores and low-mass protostars, high-mass IR-dark clump, high-mass protostellar objects and comets. For reference, the cosmic D/H ratio \citep[dotted-dashed black line;][]{oliveira2003} and the D/H measured in meteorites (red shadow) and proto-solar nebula (dotted cyan line) are reported. }
    \label{fig:DoverHlit}
\end{figure}

Using the method described in Section~\ref{sec:abundances}, we derive D/H ratios ranging from 0.01 to 0.09, all enhanced by several orders of magnitude with respect to the typical cosmic D/H ratio $\sim$10$^{-5}$ \citep{oliveira2003}. The deuterium fractionation values derived from DCO$^+$ and DNC are consistent within the uncertainties, while the D/H ratio from DCN is nearly a factor of three lower. This agrees with previous studies suggesting that D/H ratios inferred from DCN may be anti-correlated with other deuterated species i.e., it tends to be higher in more evolved sources, whereas the D/H ratio from other molecules is expected to peak at earlier phases \citep{fontani2015}. The D/H values reported here are approximately a factor of 2–3 lower than that derived from N$_2$D$^+$ \citep[$\sim$0.1][]{cosentino2025a}. This likely reflects the fact that, contrary to N$_2$D$^+$, other species may involve more complex formation pathways or be more sensitive to grain-surface chemistry \citep{Ceccarelli2014}.\\
\noindent 
In Figure~\ref{fig:DoverHlit}, we compare our measurements with literature values for low-mass starless/pre-stellar cores \citep{Hirota2001,Redaelli2019,Gelder2021,Petrashkevich2024,Ambrose2024} and low-mass protostars \citep{Hirota2001,jorgensen2004,emprechtinger2009,Imai2018,Gelder2021}, high-mass IR-dark clump \citep{gerner2015,Kai2024}, high-mass protostellar objects \citep{gerner2015,Gelder2022,Boucasse2022} and comets \citep{Meier1998,Biver2019, Despois2005, Crovisier2004,Drozdovskaya2020}. We find that the D/H ratios derived from DCO$^+$, DNC, and CH$_2$DOH broadly agree with those reported for low-mass starless/prestellar cores and protostars. However, the lack of compact continuum sources at 1 and 3 mm \citep{rathborne2006,barnes2021} as well as the lack of point-like IR emission \citep{paron2009,barnes2021} seems to indicate that the Clump is at an early evolutionary stage, that is either pre-stellar or very early protostellar stage. On the other hand, the D/H ratio derived from DCN in the literature shows little variation across different environments. DCN lines have relatively lower S/N compared with other D-bearing tracers. Hence, the D/H measurements from DCN are usually less constrained. It is therefore more challenging to establish significant variations even considering the large samples extracted from the literature. In Figure~\ref{fig:DoverHlit}, we also compare our results with typical D/H ratios reported for well-known comets \citep{Meier1998,Biver2019,Despois2005,Crovisier2004,Drozdovskaya2020}, meteorites \citep{Busemann2006,Remusat2006,Robert2003, Alexander2007,Alexander2010,Alexander2012,Alexander2017}, and the protosolar nebula \citep{Lellouch2001}. The latter value is derived from measurements of Jupiter. Our values are consistent within the uncertainties with those reported for comets, although these are often upper limits. On the other hand, our values are systematically higher by at least one order of magnitude than the protosolar nebula and meteorites values. We speculate that this discrepancy may be due to the fact that multiple chemical resets may occur between the shock-induced star formation and the final planetary system stage \citep{RomeroMirza2022,Oberg2023}. In light of all this, the Clump is most likely consistent with a pre-stellar core-like deuterium chemistry. 

\subsection{Complex Organic Molecules}
\begin{figure*}
    \centering
    \includegraphics[width=\linewidth]{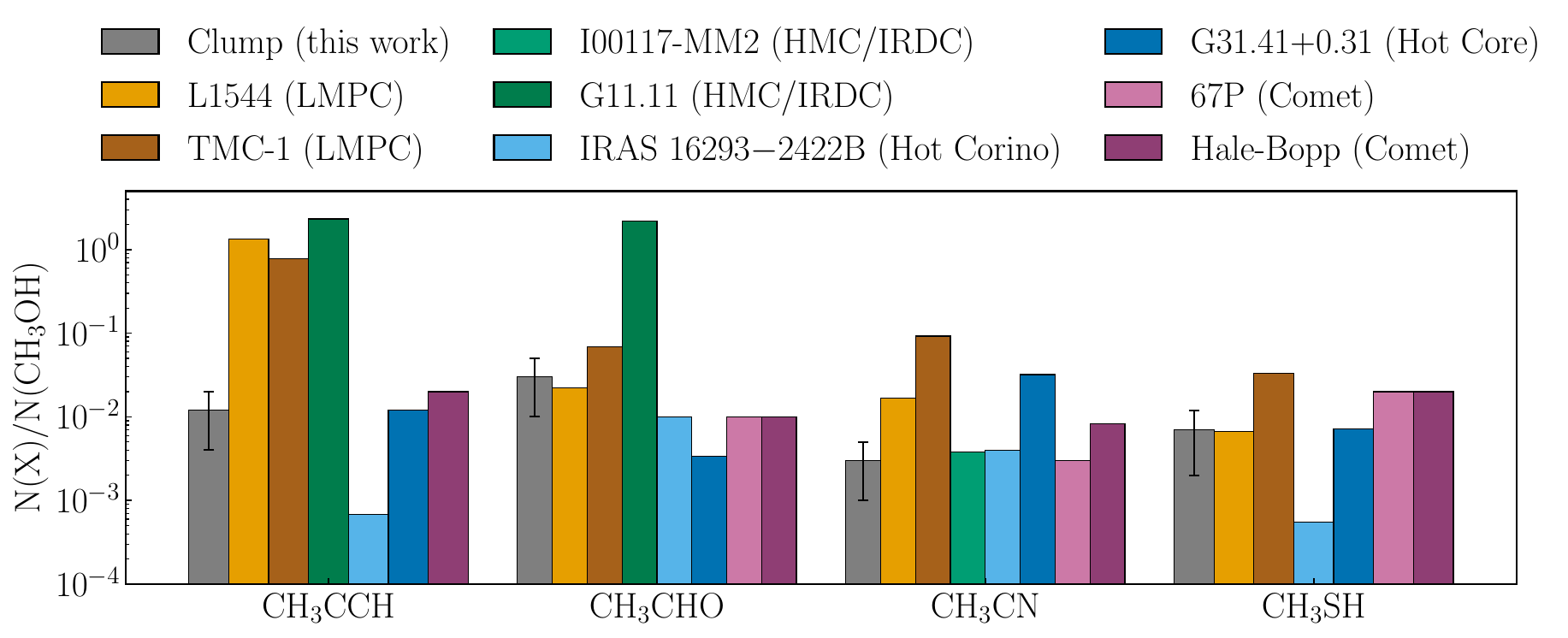}
    \caption{Comparison between COM abundances with respect to CH$_3$OH measured toward the Clump and  literature values toward L1544 \citep[LMPSC;][]{Vastel2014,Bizzocchi2014,jimenezserra2016,Vastel2019}, TMC-1 \citep[dark clouds][]{Gratier2016,Cabezas2021,Agundez2025}, I00117-MM2 and G11.11 \citep[high-mass cores and/or IRDCs][]{Vasyunina2014,fontani2015,Mininni2021}, IRAS~16293B (hot corino) and G32.41+0.31 \citep[hot core;][]{Mininni2023,LopezGallifa2024}, and the comets 67P/Churyumov–Gerasimenko and Hale–Bopp \citep{BockeleeMorvan2000,LeRoy2015,Altwegg2017}. Colors are representative of the environment: orange and dark orange for low-mass cores and dark clouds, green and dark green for massive cores/IRDCs, blue and dark blue for more evolved hot cores and corinos, pink and purple for comets.}
    \label{fig:COMsLit}
\end{figure*}
Within our high-sensitivity, broad bandwidth spectra, we have detected several complex molecules, i.e., $^{13}$CH$_3$OH, CH$_3$CCH, CH$_3$CHO, CH$_3$CN and CH$_3$SH. This is a relatively low level of complexity compared with other typical low-mass starless and pre-stellar cores \citep{jimenezserra2016,bianchi2017,Gorai2024}. 
This reduced complexity may be due to G034.77 and the Clump being located further away (several kpc) than typical low-mass sources (few hundreds of pc) and/or to the poorer sensitivity of our observations. Alternatively, the lower degree of chemical complexity may be indicative of the Clump being at an earlier evolutionary stage, similarly to what was reported by \cite{megias2023} for other low-mass starless cores. Alternatively, beam dilution effects may prevent us from detecting more complex, fainter species. Future high-angular resolution images will allow more precise estimates and to discern between these possibilities.\\
Although detections of COMs in the ISM are common, information from literature is more sparse than what was presented in Section~\ref{sec:deuterated} for D-bearing molecules, and a similar comparison cannot be carried out. Therefore, we have considered, for each broad evolutionary stage and mass regime, sources that are often regarded as an archetype of that specific environment and compared COMs abundances with those estimated toward the Clump. We have considered L1544 \citep{Vastel2014,Bizzocchi2014,jimenezserra2016,Vastel2019,jimenezserra2021} and TMC-1 \citep{Gratier2016,Cabezas2021,Agundez2025} as representative of low-mass prestellar cores (LMPSCs), I00117-MM2 and G11.11 as typical high-mass prestellar core candidates and/or IRDC environment \citep{Vasyunina2014,fontani2015,Mininni2021}, IRAS~16293B and G31.41+0.31 represent typical corinos and hot cores, respectively \citep{Kolesnikova2014,Brinkmann2020,Feng2015,Mininni2023,LopezGallifa2024,Gorai2024}. Finally, we have considered the comets 67P/Churyumov–Gerasimenko and Hale–Bopp \citep{BockeleeMorvan2000,LeRoy2015,Altwegg2017}. The comparison is presented in Figure~\ref{fig:COMsLit} where COMs abundances have been reported with respect to methanol.\\
\noindent 
From Figure~\ref{fig:COMsLit}, all COM abundances normalised to methanol are consistent with the typical cometary environment, similarly to the D-bearing species analysed in this work. The CH$_3$CCH abundance normalised to methanol is significantly lower than that observed in typical dark sources but higher than or at most similar to that expected in warmer sources. CH$_3$CCH is known to be a reliable tracer of gas excitation conditions \citep[e.g. ][]{Taniguchi2018, calcutt2019}. Although its formation pathways are highly uncertain, this species has been detected in a plethora of environments, including external galaxies \citep{Mauersberger1991, Qiu2020}. While CH$_3$CCH is often regarded as a tracer of intermediate temperatures in high-mass star forming cores \citep{fontani2003}, it can be found also in hot cores and hot corinos, both associated with the cold envelope (T$<$60 K) and a hotter, inner, component \citep[T$>$100 K;][]{HerbstDishoeck2009}. Toward the Clump, we report very low T$_{\rm{ex}}\sim$7 K. At these low values of excitation temperature, the kinetic temperature is expected to be similar \citep{Walmsley1983,zeng2018}, indicating that the CH$_3$CCH emission is associated with cold gas, i.e., not compatible with hot core/corinos conditions as seen in Figure~\ref{fig:COMsLit}. This is also supported by the lack of IR point-like source within the beam. However, from Figure~\ref{fig:COMsLit}, the CH$_3$CCH abundance with respect to CH$_3$OH is also slightly lower than what was reported toward L1544 and TMC-1. A possible explanation is that the emission from these species is unresolved in our single pointing data and may have different spatial extent. This is instead not the case for L1544 and TMC-1 which are closer and have been studied in much more detail, leading to more accurate measurements. Moreover, while the Clump is not spatially coincident with the shock front detected in \cite{cosentino2018,cosentino2019}, it is possible that some contamination is present within the beam. Higher-angular resolution images will shed light on these scenarios.\\
\noindent
Contrary to CH$_3$CCH, the species CH$_3$CHO shows abundance with respect to methanol consistent with that of typical LMPSCs and slightly higher than that observed in warmer regions. Indeed, the low excitation temperature measured for this species, T$_{\rm{ex}}\sim$5 K, suggests that the emission may be sub-thermally excited. Moreover, the ratio with respect to CH$_3$OH reported here, $\sim$0.03, is more than one order of magnitude larger than that typically measured in shocks from molecular outflows \citep[10$^{-4}$-10$^{-3}$][]{holdship2019,codella2020,lopezSepulcre2024}.\\
\noindent 
CH$_3$CN, as seen in Figure~\ref{fig:COMsLit}, shows an abundance with respect to methanol significantly lower than that measured in LMPSCs \citep[e.g.][]{Bhat2023} and consistent with that typically measured in hot corinos but not towards hot cores. Similarly to CH$_3$CCH, this may be due to the fact that the true emission spatial extent remains unresolved in our observations leading to beam-dilution effects at the larger source distance. CH$_3$CN has been widely used to investigate the physical conditions of warm environments, and \cite{codella2009} showed that it can be efficiently produced also in shocked regions. While the excitation temperature measured here is lower than that reported by \cite{codella2009}, the abundance with respect to methanol is consistent with that reported toward the shocked knots of the L1157 outflows \citep[see also][]{Arce2008}. %Therefore, we cannot exclude the possibility that CH$_3$CN may be associated with shocked gas whose emission is contaminating the relatively large beam of our observations. In addition to this, \cite{cosentino2025a} have discussed the possibility that an older shock, non detected in SiO, might have already compressed the region. Hence, CH$_3$CN may not be associated with the Clump, but rather be a residual of older shocked material.\\

From Figure~\ref{fig:COMsLit}, the CH$_3$SH abundance w.r.t. methanol is consistent with that estimated toward low-mass starless/prestellar cores and hot cores. Toward LMPSCs, it has been suggested that CH$_3$SH forms onto the icy mantle of dust grains and it is then released into the gas phase through non-thermal processes, such as cosmic ray-induced desorption \citep[e.g.][]{Zhou2022}. However, the formation pathway of these species is still unclear and follow-up high-resolution observations are needed to spatially resolve the emission and further constrain the emitting regions for CH$_3$SH and CH$_3$CN. 

\section{Summary and Conclusions}\label{sec:conclusions}
We have presented high-sensitivity single pointing observations obtained at 3 mm (IRAM 30m) and 7 mm (Yebes 40m) toward the shock-induced potentially prestellar cluster known as the Clump \citep{cosentino2023}. Within the large bandwidth spectra, we have identified multiple D-bearing species such as DCO$^+$, DNC, DCN and CH$_2$DOH as well as multiple COMs, i.e., CH$_3$OH, CH$_3$CHO, CH$_3$CCH, CH$_3$CN and CH$_3$SH. The D/H ratios measured from the detected D-bearing species and their H-bearing counterparts are several order of magnitude larger than the typical cosmic D/H ratio and broadly consistent with those observed in low-mass starless/pre-stellar cores. %The only exception being the D/H measured from DCN, expected to be lower in environments at earlier stages of evolution. In addition, we report the detection of deuterated methanol, so far only detected in few nearby cold and dark sources.\\
Within the high-sensitivity spectra, we also detected multiple COMs. The Clump shows a level of chemical complexity similar to that reported toward some young low-mass prestellar cores (younger than L1544) such as  L1517B \citep{megias2023}. This may be indicative of the source evolutionary stage, as already reported toward other low-mass star starless cores. Alternatively, this may be a consequence of beam dilution effects. Among the detected species, CH$_3$CHO and CH$_3$SH show abundance w.r.t. methanol consistent with that observed in low-mass starless/pre-stellar cores. On the other hand, the CH$_3$CCH and CH$_3$CN abundance normalised to methanol is slightly lower than values reported in low-mass environments, suggesting potential contamination from shocked gas. 
The chemical complexity and Deuterium content of the region are consistent with the shock time-scales estimated in \cite{cosentino2019}. Indeed, the enhanced densities reported for the post-shocked gas shorten the CO depletion time-scale, boosting the production of deuterated species \citep{lis2002,lis2016}. For the post-shock density estimated by \cite{cosentino2019}, the CO depletion timescale is expected to be $\leq$10$^4$ yrs \citep{caselli1999}, consistent with both the shock dynamical age \citep[few 10$^4$ yr;][]{cosentino2019} and the SNR age (20 kyr). Finally, we have compared the abundances of all detected D-bearing species and COMs with those observed in comets and found a broad agreement. This similarity suggests that COM and Deuterium budgets could be inherited from the compressed, cold and dense post-shocked material, where the shock passage has enabled a fast chemistry. In light of all this, we suggest that the Clump is a shock-induced dense core at the very early stage of evolution with the potential to undergo gravitational collapse and form the next generation of protostars in the cloud. We also speculate that stellar-feedback slow-shocks may contribute to set the initial chemical conditions of star formation, which is later on inherited by comets. Follow-up high-angular resolution images are needed to remove potential beam dilution effect and separate the emission arising from cores and the shocked gas. Nevertheless, at the best of our knowledge, this represents the first detection of COMs toward a SNR-molecular cloud interaction site, and it is a potentially unique signature of chemically enriched, post-shock dense core formation. Finally, shocks from late-stage SNRs are intrinsically gentler and more spatially extended than than those observed in molecular outflows. Hence, if some of the detected COMs have been released into the gas phase by the SNR-shock passage, this may lead the way to new investigation into COMs survival in shocked environments. 

\begin{acknowledgements}
This work is based on observations carried out under project number 027-21 with the IRAM 30 m telescope. IRAM is supported by INSU/CNRS (France), MPG (Germany) and IGN (Spain). 
This work is also based on observations carried out with the Yebes 40m telescope (project 24A019).The 40m radio telescope at Yebes Observatory is operated by the Spanish Geographic Institute (IGN; Ministerio de Transportes, Movilidad y Agenda Urbana). J.C.T. acknowledges support from
ERC project 788829–MSTAR. I.J-.S acknowledges funding from grant PID2022-136814NB-I00 funded by the Spanish Ministry of Science, Innovation and Universities/State Agency of Research MICIU/AEI/ 10.13039/501100011033 and by “ERDF/EU”. R.F. acknowledges support from the grants Juan de la Cierva FJC2021-046802-I, PID2020-114461GB-I00 and
CEX2021-001131-S funded by MCIN/AEI/ 10.13039/501100011033 and by “European Union NextGenerationEU/PRTR”. S.V. acknowledges partial funding from the European Research Council (ERC) Advanced Grant MOPPEX 833460.      
\end{acknowledgements}


\bibliographystyle{aa} % style aa.bst
\bibliography{aa.bib}

\begin{appendix}
\section{Spectra of detected species}
\begin{figure*}[!b]
    \centering
    \includegraphics[width=0.9\textwidth,trim=0cm 20cm 0cm 0cm, clip=True]{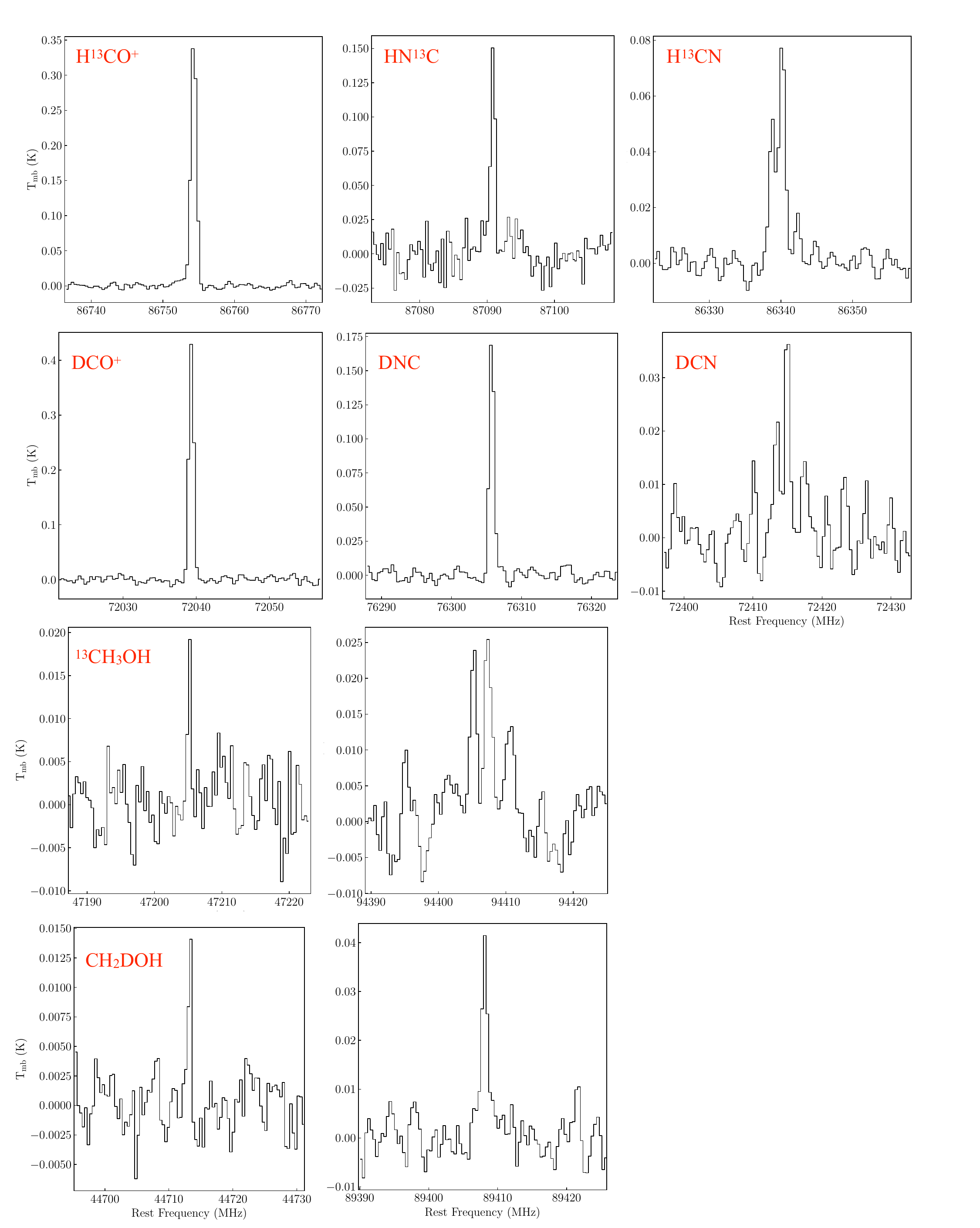}
    \caption{Spectra of detected D-bearing species and their H-bearing counterparts. For clarity, we only indicate the species, while quantum numbers for each transitions are reported in Table~\ref{tab:LineList}.}
    \label{fig:spec1}
\end{figure*}

\begin{figure*}
    \centering
    \includegraphics[width=0.85\linewidth,trim=0cm 10cm 0cm 0cm, clip=True]{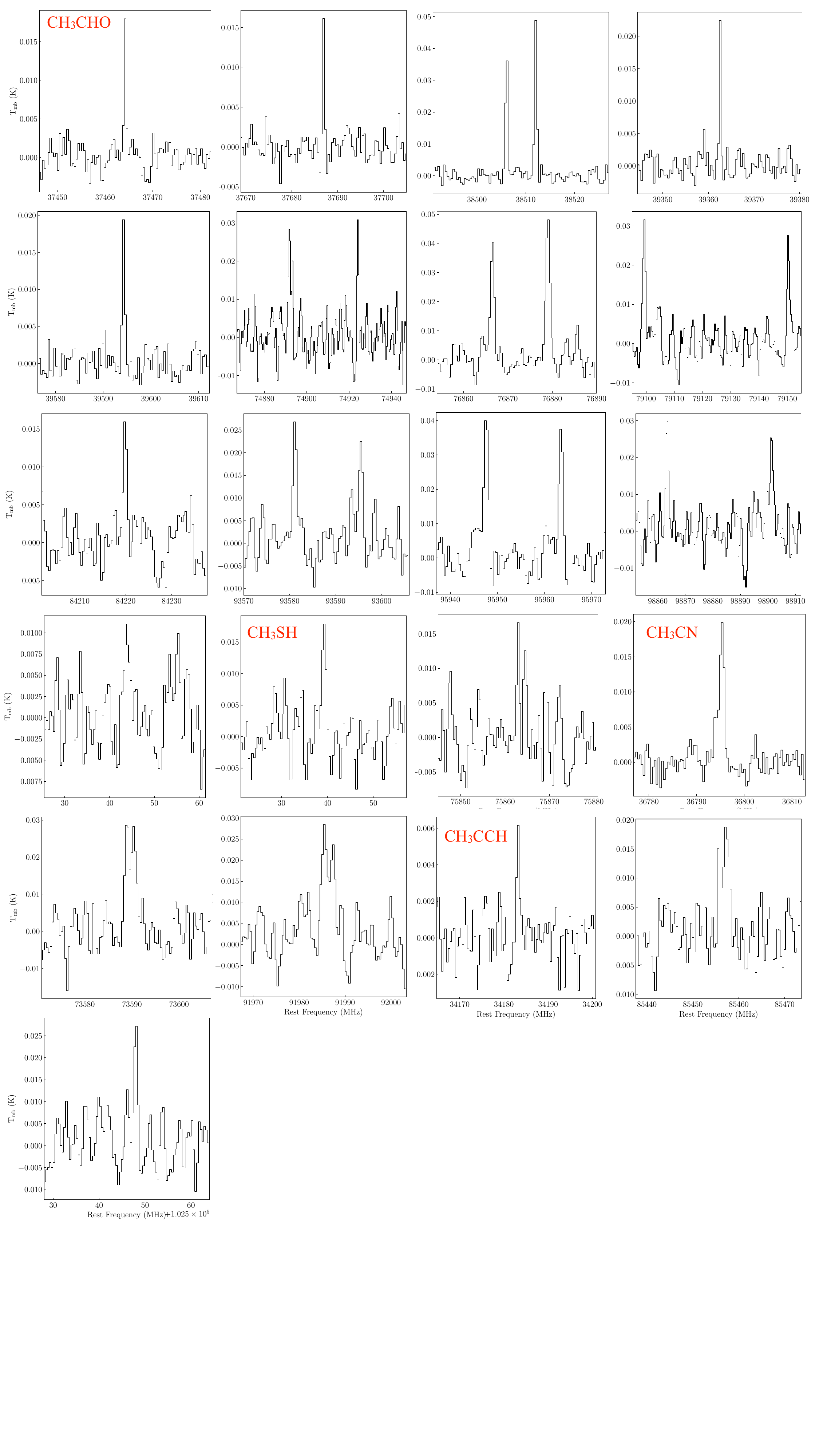}
    \caption{Spectra of detected COMs. For clarity, we only indicate the species, while quantum numbers for each transitions are reported in Table~\ref{tab:LineList}.}
    \label{fig:spec2}
\end{figure*}
\end{appendix}

\end{document}